\input{aipcheck}

\documentclass[
    ,final            % use final for the camera ready runs
  ]
  {aipproc}

\layoutstyle{6x9}

\newcommand{\beq}{\begin{equation}}
\newcommand{\eeq}{\end{equation}}
\newcommand{\gev}{{\rm GeV}}
\def\su{{\rm SUSY}}

\begin{document}

\title{QCD Effects in Cosmology}


\author{Jose A. R. Cembranos}{
  address={William I. Fine Theoretical Physics Institute,
University of Minnesota, Minneapolis, 55455, USA \\
${^2}$ School of Physics and Astronomy,
University of Minnesota, Minneapolis, 55455, USA}
}

\begin{abstract}
The cosmological evolution in the radiation dominated regimen is usually computed by assuming an ideal relativistic thermal bath. In this note, we discuss the deviation from the non-interaction assumption. In either the standard model (SM) and the minimal supersymmetric standard model (MSSM), the main contribution comes from the strong interaction. An understanding of these effects are important for precision measurements and for the evolution of scalar modes, where the commented corrections constitute the main source of the dynamics.
\end{abstract}

\maketitle

%%%%%%%%%%%%%%%%%%%%%%%%%%%%%%%%%%%%%%%%%%%%
%% MAINMATTER
%%%%%%%%%%%%%%%%%%%%%%%%%%%%%%%%%%%%%%%%%%%%

\section{The Relativistic Thermal Bath}
\label{thermo}

\begin{figure}[h]
\centerline{
\includegraphics[width=0.43\textwidth,angle=0]{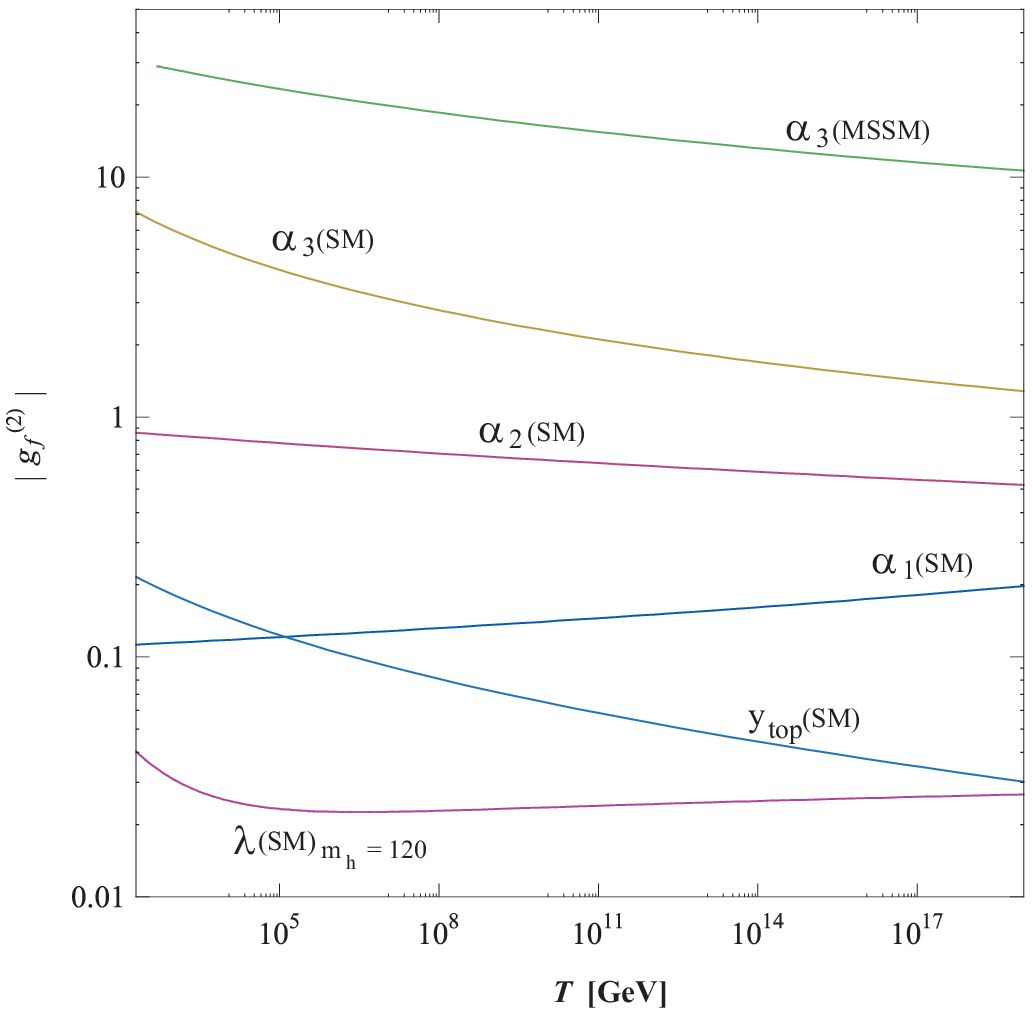}
\hspace{1cm}
\includegraphics[width=0.43\textwidth,angle=0]{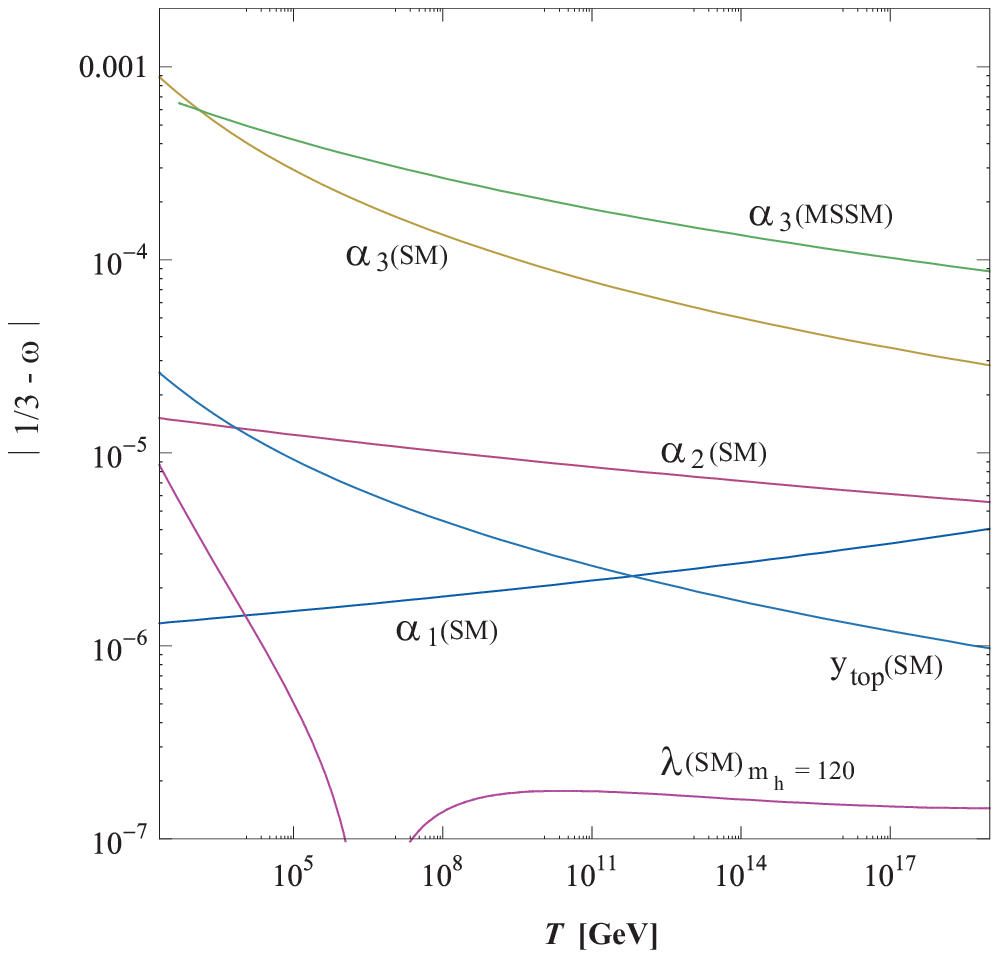}
}
\caption{Left panel: The absolute value of the leading corrections (two-loops) to the effective free energy number of relativistic
degrees of freedom from the different gauge interactions, top Yukawa and Higgs self-coupling in the SM.  The strong contribution 
($\alpha_3$), is dominant and is also plotted for the case of the MSSM assuming that all supersymmetric partners
have masses fixed at $m_{\su} = 500 \; \gev$. All contributions to $g_{f}^{(2)}$ 
are negative.
Right panel: The absolute value of the leading correction to $1/3-\omega$ from the same interactions shown in the left panel. The strong 
contribution is dominant  not only due to the fact that the correction to $g_{f}^{(2)}$ is bigger, but also because 
the running of $\alpha_3$ is steeper. On the other hand, the contribution from the hypercharge interaction (and the Higgs coupling 
for high temperatures) is negative and opposite to the others since $\alpha_1$ ($\lambda$) increases with $T$. 
Read \cite{copu} for more details.
}
\label{fig:w-sm-mssm}
\end{figure}

We can describe the thermodynamical properties of any fluid by its free energy density (${\cal F}$). Equivalently, we can use $g_f$, the {\it effective free energy number of relativistic degrees of freedom}, that is defined by normalizing to the free energy density of one non-interacting massless bosonic particle \cite{copu}. The partition function and $g_f$ can be computed diagrammatically \cite{Kapusta:1989tk}. The one loop vacuum diagrams provide the free theory result. In this case, the coefficient $g_f$ is simply the number of bosonic ($N_b$) and fermionic ($N_f$) relativistic degrees of freedom: $g_{f,{\rm free}}= N_b+7\, N_f/8$, where the $7/8$ coefficient multiplying the fermionic contribution is due to the different value of the partition function obtained from the Fermi-Dirac thermal distribution rather than the Bose-Einstein one. Higher loops account for the interactions. In \cite{copu}, it has been shown that the strong interaction is dominant even at very high temperatures (see Fig. \ref{fig:w-sm-mssm}) except for the case of a heavy Higgs boson. Unfortunately, the situation is not simple since the finite temperature perturbation approach to QCD has a very slow convergence \cite{Kajantie:2002wa}. In any case, the leading order correction coming from the strong interaction has the right order and it can be used to estimate the interaction effects in the thermal bath (read \cite{copu} for a particular discussion inside the SM and the MSSM)

In the right panel of Fig. \ref{fig:w-sm-mssm} we show different contributions to the departure of the equation of state from the $w = 1/3$ value corresponding to the non-interacting case. The effect decreases as the temperature (T) increases, since  $\alpha_3$ decreases with T (and, consequently the absolute value of the two loop contribution to $g_f$ decreases). At the lowest temperatures shown, the departure is stronger in the SM than in the MSSM case. This is because $\alpha_3$ runs faster in the SM. However, since $\alpha_3$ decreases less in the MSSM, the departure from the non-interacting value becomes greater for the MSSM as the temperature increases (see Fig. \ref{fig:w-sm-mssm} and \cite{copu}).

\subsection{Precision Cosmology}

Interaction effects in the relativistic thermal bath are important in the precision era of astrophysical observations that we have already entered. 
We can illustrate this effect with the thermal relic abundance. Although there are other possibilities \cite{other}, Dark Matter (DM) is usually assumed to be in the form of stable Weakly-interacting massive particles (WIMPs) that naturally freeze-out with the right thermal abundance. WIMPs emerge in different well-motivated particle physics scenarios as in R-parity conserving supersymmetry (SUSY) models~\cite{SUSY1,SUSY2}, universal extra dimensions (UED)~\cite{UED1, UED2}, or brane-worlds~\cite{BW1,BW2,BW3}.

The present uncertainty on the total DM density using data from the {\it Wilkinson Microwave Anisotropy Probe} (WMAP) is around $6\%$ (five years data \cite{Dunkley:2008ie}). This precision can be improved to $3\%$ when measurements of Baryon Acoustic Oscillations and Type Ia supernovae are also taking into account \cite{Komatsu:2008hk}. On the other hand, the prospects for the precision that can be obtained by analyzing data from the {\it Planck Surveyor}, reduce this uncertainty even below $1\%$. However, in order to associate the observed precision with the thermal relic density of a particular WIMP model, it is neccesary to understand the thermal bath up to the same level. Unfortunately, the uncertainties are bigger than $1\%$ percent even at high temperatures.  The situation is even worse because typical WIMPs freeze out at a temperature of $10$ GeV approximately, i.e. much closer to the QCD phase transition. In this case, it is clear that the perturbative approach does not converge and non-perturbative analises (such as lattice studies) are necessary. Alternatively, if we are able to understand the nature of DM from other experiments as the new generation of colliders \cite{Coll}, we may have the oportunity to improve our knowledge about QCD by making precise astrophysical observations. 

\subsection{Scalar Dynamics}

On the other hand, there is a case in which the deviation from the ideal gas is not only important but also fundamental, since it constitutes the leading order term of the dynamics. It happens for the case of scalar fields that couple to the trace of the energy momentum tensor. These scalars are ubiquitous in many theories, as for example, Jordan-Fierz-Brans-Dicke (JFBD) scalars in Scalar-tensor theories of gravity such as f(R) theories \cite{stgen}, moduli fields from string theories such as the dilaton \cite{polchy}, or graviscalars from extra dimensional models such as the radion 
\cite{gravrad}.

The dynamics of these theories is usually described in what is called the Jordan Frame or Einstein Frame. Inside the Jordan Frame, the standard matter is coupled minimally to the Jordan metric. It implies that the particle masses are constant, whereas the Planck scale depends on the particular value of the scalar field. Alternatively, it is possible to use the Einstein Frame, where the action for the metric is the standard, i.e. it is described by the Einstein-Hilbert action, but the matter content is coupled explicitly to the scalar mode. In this frame, the Planck scale is constant but the particle masses depend on the value of the scalar field.

Both descriptions are classically equivalent and are related by a simple conformal transformation. The cosmological evolution of the scalar field is determined by the standard equation of motion of a scalar field with an extra source term proportional to the trace of the energy momentum tensor. If the relativistic thermal bath is ideal, this term is zero. It means that the scalar field is fundamentally frozen at its initial value.

The situation is completely different if this source term is non-zero, even if it is small. The conformal factor may evolve orders of magnitude depending on the particular coupling. For example, in \cite{copu}, it has been studied a model with quadratic coupling and without potential. In this case, the coupling works as an attractor to the general relativity limit. This important effect implies that the early cosmology can be very different from standard. For example, depending of initial conditions, there is a maximum in the Jordan temperature that can solve or alleviate the unwanted relic problem since it could avoid the production of dangerous solitons or unstable gravitinos \cite{copu}.

\vspace{0.5cm}
In conclusion, we have discussed briefly the deviations from the ideal relativistic thermal bath, with interesting consequences for precision measurements and the dynamics of scalar modes.

\begin{theacknowledgments}
This work is supported in part by DOE grant DOE/DE-FG02-94ER40823, FPA 2005-02327 (DGICYT, Spain), and CAM/UCM 910309 projects.
\end{theacknowledgments}

\bibliographystyle{aipproc}   % if natbib is available

\end{document}